\documentstyle[12pt]{article}

\def\a{\alpha}
\def\b{\beta}
\def\m{\mu}
\def\n{\nu}
\def\tm{\tilde{m}}
\def\th{\theta}
\def\tr{{\rm Tr}}
\def\pf{{\rm Pf}}
\def\g{\gamma}
\def\de{\delta}
\def\P{\Phi}

\def\l{\lambda}
\def\L{\Lambda}
\def\s{\sigma}

\def\e{\epsilon}

\def\t{\tau}
\def\G{\Gamma}
\def\Z{{\bf Z}}
\def\Bb{\bar{B}}
\def\Qb{\bar{Q}}
\def\ib{\bar{i}}
\def\jb{\bar{j}}
\def\implies{\Rightarrow}

\begin{document}

\renewcommand{\theequation}{\thesection.\arabic{equation}}
\newcommand{\beq}{\begin{equation}}
\newcommand{\eeq}[1]{\label{#1}\end{equation}}
\newcommand{\ber}{\begin{eqnarray}}
\newcommand{\eer}[1]{\label{#1}\end{eqnarray}}

\begin{titlepage}
\begin{center}

                                 \hfill    CERN-TH/96-268 \\
                                 \hfill    hep-th/9611152 \\

\vskip .5in

{\large \bf ON NON-PERTURBATIVE RESULTS \\
            IN SUPERSYMMETRIC GAUGE THEORIES \\
            -- A LECTURE}\footnotemark \\
\footnotetext{Lecture presented at the Workshop on {\em Gauge
Theories, Applied Supersymmetry and Quantum Gravity}, Imperial College,
London, UK, 5-10 July, 1996. To appear in the proceedings.}

\vskip .4in

{\large  Amit Giveon}\footnotemark \\
\footnotetext{Permanent address: Racah Institute of Physics, The
Hebrew University, Jerusalem 91904, Israel; e-mail address:
giveon@vms.huji.ac.il}
\vskip .1in
{\em  Theory Division, CERN, CH-1211, Geneva 23, Switzerland} \\
\vskip .15in

\end{center}
\vskip .2in
\begin{center} {\bf ABSTRACT } \end{center}
\begin{quotation}
\noindent
Some notions in non-perturbative dynamics of supersymmetric gauge
theories are being reviewed. This is done by touring through a few examples.
\end{quotation}

\vfill

\begin{flushleft}
CERN-TH/96-268 \\
September 1996
\end{flushleft}

\end{titlepage}
\eject
\def\baselinestretch{1.2}
\baselineskip 16 pt


\section{Introduction}
\setcounter{equation}{0}

In this lecture, we present some notions in supersymmetric Yang-Mills (YM)
theories. We do it by touring through a few examples where we face a variety of
non-perturbative physics effects -- infra-red (IR) dynamics of gauge theories.

We shall start with a general review; some of the points we consider follow the
beautiful lecture notes in~\cite{lec}. \\

\begin{center}
\bf
Phases of Gauge Theories
\end{center}
\vskip .1in

There are three known phases of gauge theories:
\begin{itemize}
\item
{\em Coulomb Phase}: there are massless vector bosons (massless photons $\g$;
no confinement of both electric and magnetic charges). The behavior of the
potential $V(R)$ between
electric test charges, separated by a large distance $R$, is $V(R)\sim 1/R$;
the electric charge at large distance behaves like a constant: $e^2(R)\sim
constant$. The potential of magnetic test charges separated by a large
distance behaves like $V(R)\sim 1/R$, and the magnetic charge behaves like
$m^2(R)\sim constant$, $e(R)m(R)\sim 1$ (the Dirac condition).
\item
{\em Higgs Phase}: there are massive vector bosons ($W$ bosons and $Z$ bosons),
electric charges are condensed (screened) and magnetic charges are confined
(the Meissner effect). The potential between magnetic test charges separated by
a large distance is $V(R)\sim \rho R$ (the magnetic flux is confined into a
thin tube, leading to this linear potential with a string tension $\rho$). The
potential between electric test charges is the Yukawa potential; at large
distances $R$ it behaves like a constant: $V(R)\sim constant$.
\item
{\em Confining Phase}: magnetic charges are condensed (screened) and electric
charges are confined. The potential between electric test charges separated by
a large distance is $V(R)\sim \s R$ (the electric flux is confined into a thin
tube, leading to the linear potential with a string tension $\s$).
The potential
between magnetic test charges behaves like a constant at large distance $R$.
\end{itemize}

\begin{center}
\bf
Remarks
\end{center}

\begin{enumerate}
\item
In addition to the familiar Abelian Coulomb phase, there are theories which
have a {\em non-Abelian Coulomb phase}~\cite{S2}, namely, a theory with
massless
interacting quarks and gluons exhibiting the Coulomb potential. This phase
occurs when there is a non-trivial IR fixed point of the renormalization group.
Such theories are part of other possible cases of non-trivial, {\em
interacting $4d$ superconformal field theories} (SCFTs)~\cite{AD,apsw}.
\item
When there are matter fields in the fundamental representation of the gauge
group, virtual pairs can be created from the vacuum and screen the sources. In
this situation, there is no invariant distinction between the Higgs and the
confining phases~\cite{BR}. In particular, there is no phase with a potential
behaving as $V(R)\sim R$ at large distance, because the flux tube can break.
For large VEVs of the fields, a Higgs description is most natural, while for
small VEVs it is more natural to interpret the theory as ``confining.''
It is possible to smoothly interpolate from one interpretation to the other.
\item
{\em Electric-Magnetic Duality}: Maxwell theory is invariant under
\beq
{\bf E}\to {\bf B}, \qquad {\bf B}\to -{\bf E},
\eeq{EB}
if we introduce magnetic charge $m=2\pi/e$ and also interchange
\beq
e\to m, \qquad m\to -e .
\eeq{em}
Similarly, Mandelstam and `t Hooft suggested that under electric-magnetic
duality the Higgs phase is interchanged with a confining phase. Confinement
can then be understood as the dual Meissner effect associated with a condensate
of monopoles.

Dualizing a theory in the Coulomb phase, one remains in the same phase. For
an
Abelian Coulomb phase with massless photons, this electric-magnetic duality
follows from a standard duality transformation, and is extended to $SL(2,\Z)$
S-duality, acting on the complex gauge coupling by
\beq
\t\to {a\t+b\over c\t+d}, \qquad \t={\theta\over 2\pi}+i{4\pi\over g^2}, \qquad
a,b,c,d\in\Z, \quad ad-bc=1 .
\eeq{tabcd}
In the non-Abelian Coulomb phase this duality is not obvious. The simplest
versions of such S-duality in non-trivial interacting theories are in $N=4$
supersymmetric YM theory~\cite{olive}, and in finite $N=2$ supersymmetric YM
theories~\cite{SW2}. Electric-magnetic duality can be extended, sometimes, to
asymptotically free $N=1$ theories~\cite{lec}.
\end{enumerate}

\section{The Low-Energy Effective Theory}
\setcounter{equation}{0}
We will consider the low-energy effective action for the light fields,
${\cal L}_{eff}$(light fields), integrating out degrees of freedom
(such as massive vector bosons, resonances, etc.) above some scale.
${\cal L}_{eff}$ has a linearly realized supersymmetry (as long as we are above
the scale of possible supersymmetry breaking).

Supersymmetry can be made manifest by working with superfields~\cite{books}.
All renormalizable Lagrangians can be constructed in terms of chairal
(= scalar), anti-chiral and vector supermultiplets:
\begin{itemize}
\item
{\em Chiral and Anti-Chiral Superfields}: the light matter fields can be
combined into chiral and anti-chiral superfields
\beq
X_r=\phi_r+\theta_{\a}\psi_r^{\a}+... , \qquad
X_r^{\dag}=\phi_r^{\dag}+\bar{\theta}_{\dot\a}\psi^{\dag\dot\a}_r+...\, .
\eeq{XXdag}
Chiral (anti-chiral) superfields obey ${\bar D}X_r=0$ ($DX_r^{\dag}=0$).
\item
{\em Vector Supermultiplet}: the real superfield $V=V^{\dag}$
combining the light vector bosons $A_{\m}$ and the gauginos $\l_{\a}$,
$\l_{\dot\a}^{\dag}$; schematically,
\beq
V\sim
\theta\s^{\m}\bar{\theta}A_{\m}+\theta^2(\bar{\theta}\l^{\dag})
+\bar{\theta}^2(\theta\l)+...\, .
\eeq{Vector}
\end{itemize}

\noindent
By {\em effective}, we mean in Wilson sense:
\beq
\int [{\rm modes}\,\, p>\m]e^{-S}=e^{-S_{eff}(\m,{\rm light}\,\,{\rm modes})} ,
\eeq{wilson}
so, in principle, ${\cal L}_{eff}$ depends on a scale $\m$. But due to
supersymmetry, the dependence on the scale $\m$ disappear (except for the gauge
coupling $\t$ which has a $\log\m$ dependence).

When there are no interacting massless particles, the Wilsonian effective
action = the 1PI effective action; this is often the case in the Higgs or
confining phases.

\subsection{The Effective Superpotential}
We will focus on a particular contribution to  ${\cal L}_{eff}$ -- the
effective superpotential term:
\beq
{\cal L}_{int}\sim \int d^2\theta W_{eff}(X_r,g_I,\L) + c.c ,
\eeq{Wint}
where $X_r$ = light chiral superfields, $g_I$ = various coupling constants, and
$\L$ = dynamically generated scale (associated with the gauge dynamics):
$\log (\L/\m)\sim -8\pi^2/g^2(\m)$. Integrating over $\theta$, the
superpotential gives a scalar potential and Yukawa-type interaction of scalars
with fermions.

The quantum, effective superpotential $W_{eff}(X_r,g_I,\L)$ is constrained by
holomorphy, global symmetries and various limits~\cite{seib,lec}:
\begin{enumerate}
\item
{\em Holomorphy}: supersymmetry requires that $W_{eff}$ is holomorphic in the
chiral superfields $X_r$ ({\em i.e.}, independent of the $X_r^{\dag}$).
Moreover, we will think of all the coupling constants $g_I$ in the tree-level
superpotential $W_{tree}$ and the scale $\L$ as background chiral superfield
sources. This implies that $W_{eff}$ is holomorphic in $g_I$, $\L$ ({\em i.e.},
independent of $g_I^*$, $\L^*$).
\item
{\em Symmetries and Selection Rules}: by assigning transformation laws both to
the fields and to the coupling constants (which are regarded as background
chiral superfields), the theory has a large global symmetry. This implies that
$W_{eff}$ should be invariant under such global symmetries.
\item
{\em Various Limits}: $W_{eff}$ can be analyzed approximately at weak coupling,
and some other limits (like large masses).
\end{enumerate}

\noindent
Sometimes, holomorphy, symmetries and  various limits are strong enough to
determine $W_{eff}$! The results can be highly non-trivial, revealing
interesting non-perturbative dynamics.

\subsection{The Gauge ``Kinetic Term'' in a Coulomb Phase}
When there is a Coulomb phase, there is a term in ${\cal L}_{eff}$ of the form
\beq
{\cal L}_{gauge}\sim \int d^2\theta {\rm Im}\Big[\tau_{eff}(X_r,g_I,\L)
{\cal W}_{\a}^2\Big] ,
\eeq{Lgauge}
where ${\cal W}_{\a}$ = gauge supermultiplet (supersymmetric field strength);
schematically,
${\cal W}_{\a}\sim \l_{\a}+\theta_{\b}\s_{\a}^{\m\n\,\b}F_{\m\n}+...$.
Integrating over $\theta$, ${\cal W}_{\a}^2$ gives the term $F^2+iF\tilde{F}$
and its supersymmetric extension. Therefore,
\beq
\t_{eff}={\theta_{eff}\over 2\pi}+i{4\pi\over g^2_{eff}}
\eeq{teff}
is the effective, complex gauge coupling. $\tau_{eff}(X_r,g_I,\L)$ is also
holomorphic in $X_r,g_I,\L$ and, sometimes, it can be exactly determined by
using holomorphy, symmetries and various limits.

\subsection{The ``Kinetic Term''}
The kinetic term is determined by the K\" ahler potential $K$:
\beq
{\cal L}_{kin}\sim \int d^2\theta d^2\bar{\theta} K(X_r,X_r^{\dag}) .
\eeq{Lkin}
If there is an $N=2$ supersymmetry, $\t_{eff}$ and $K$ are related; for an
$N=2$ supersymmetric YM theory with a gauge group $G$ and in a Coulomb phase,
${\cal L}_{eff}$ is given in terms of a single holomorphic function
${\cal F}(A^i)$:
\ber
{\cal L}_{eff}&\sim& {\rm Im}\left[\int d^4\theta {\partial {\cal F}\over
\partial X^i}X^{\dag i}+{1\over 2}\int d^2\theta {\partial^2  {\cal F}\over
\partial X^i \partial X^j} {\cal W}^{\a i}{\cal W}_{\a}^j\right],
\nonumber\\ &&  i,j=1,...,{\rm rank}G.
\eer{N2eff}
A manifestly gauge invariant $N=2$ supersymmetric action which reduces to the
above at low energies is~\cite{givroc}
\ber
&&  {\rm Im}\left[\int d^4\theta {\partial {\cal F}\over \partial  X^a}
(e^V)_{ab} X^{\dag b}+{1\over 2}\int d^2\theta {\partial^2  {\cal
F}\over \partial X^a \partial X^b} {\cal W}^{\a a}{\cal W}_{\a}^b\right],
\nonumber\\  && a,b=1,...,{\rm dim}G.
\eer{N2cov}
\\
This concludes the general review. Next, we consider some examples of results.

\section{Summary of Results in $N=1$ Supersymmetric $SU(2)$ Gauge Theories}
\setcounter{equation}{0}
In the next few sections, we shall summarize some results in $4d$, $N=1$
supersymmetric $SU(2)$ gauge theories: the exact effective superpotentials,
the vacuum structure, and the exact effective Abelian couplings
for arbitrary bare masses and Yukawa couplings. A few generalizations to other
gauge groups will be pointed out. \\

\begin{center}
\bf{The Main Result}
\end{center}
\vskip .1in

\noindent
The results in this summary are based on some of the results in
refs.~\cite{efgr,efgr2}.~\footnote{Some of these results also appear in the
proceedings~\cite{efgr-proc} of the
{\em 29th International Symposium on the Theory of Elementary Particles} in
Buckow, Germany, August 29 - September 2, 1995, and of the workshop on
{\em STU-Dualities and
Non-Perturbative Phenomena in Superstrings and Supergravity}, CERN, Geneva,
November 27 - December 1, 1995.}
We consider $N=1$ supersymmetric $SU(2)$ gauge theories in four dimensions,
with any possible content of matter superfields, such that the theory is
either one-loop asymptotic free or conformal. This allows the introduction of
$2N_f$ matter supermultiplets in the fundamental representation,
$Q_i^a$, $i=1,...,2N_f$,
$N_A$ supermultiplets in the adjoint representation,
$\P_{\a}^{ab}$, $\a=1,...,N_A$,
and $N_{3/2}$ supermultiplets in the spin 3/2 representation, $\Psi$.
Here $a,b$ are fundamental representation indices, and $\P^{ab}=\P^{ba}$
(we present $\Psi$ in a schematic form as we shall not use it much). The
numbers $N_f$, $N_A$ and $N_{3/2}$ are limited by the condition:
\beq
b_1=6-N_f-2N_A-5N_{3/2}\geq 0,
\eeq{b1}
where $-b_1$ is the one-loop coefficient of the gauge coupling beta-function.

The main result of this section is the following: the effective
superpotential of an (asymptotically free or conformal)
$N=1$ supersymmetric $SU(2)$ gauge theory,
with $2N_f$ doublets and $N_A$ triplets (and $N_{3/2}$ quartets) is
\ber
&&W_{N_f,N_A}(M,X,Z,N_{3/2}) = \nonumber\\
&&-\de_{N_{3/2},0} (4-b_1)\Big\{\L^{-b_1} \pf_{2N_f} X\Big[
{\rm det}_{N_A}(\Gamma_{\a\b})\Big]^2
\Big\}^{1/(4-b_1)}\nonumber\\
&&+\tr_{N_A} \tm M +{1\over 2}\tr_{2N_f} mX
+{1\over\sqrt{2}}\tr_{2N_f} \l^{\a} Z_{\a}+\de_{N_{3/2},1} gU ,
\eer{W}
where
\beq
\Gamma_{\a\b}(M,X,Z)=M_{\a\b}+\tr_{2N_f}(Z_{\a}X^{-1}Z_{\b}X^{-1}).
\eeq{G}
The first term in (\ref{W}) is the exact (dynamically generated)
non-perturbative superpotential~\footnote{Integrating in the ``glueball'' field
$S=-{\cal W}_{\a}^2$, whose source is $\log \L^{b_1}$, gives the
non-perturbative superpotential:
$$ W(S,M,X,Z)=S\Big[\log \Big({\L^{b_1}S^{4-b_1}\over
\pf X({\rm det}\Gamma )^2}\Big)-(4-b_1)\Big] .$$}, and the other terms
form the tree-level superpotential.
$\L$ is the dynamically generated scale, while $\tm_{\a\b}$, $m_{ij}$
and $\l^{\a}_{ij}$ are the bare masses and Yukawa couplings, respectively
($\tm_{\a\b}=\tm_{\b\a}$, $m_{ij}=-m_{ji}$,
$\l^{\a}_{ij}=\l^{\a}_{ji}$). The gauge singlets, $X$, $M$, $Z$, $U$, are
given in terms of the $N=1$ superfield doublets, $Q^a$, the triplets,
$\P^{ab}$, and the quartets, $\Psi$,  as follows:
\ber
X_{ij}&=&Q_{ia} Q_j^a, \qquad  a=1,2, \qquad i,j=1,...,2N_f,
\nonumber\\
M_{\a\b}&=&\P_{\a b}^a\P_{\b a}^b, \qquad \a ,\b=1,...,N_A, \qquad a,b=1,2,
\nonumber\\
Z_{ij}^{\a}&=&Q_{ia}\P_{\a b}^a Q_j^b , \qquad \qquad U=\Psi^4.
\eer{XMZ}
Here, the $a,b$ indices are raised and lowered with an $\e_{ab}$ tensor.
The gauge-invariant superfields $X_{ij}$ may be considered as a mixture of
$SU(2)$ ``mesons'' and ``baryons,'' while the gauge-invariant superfields
$Z_{ij}^{\a}$ may be considered as a mixture of $SU(2)$ ``meson-like'' and
``baryon-like'' operators.

Equation (\ref{W}) is a universal representation of the superpotential for all
infra-red non-trivial theories; all the physics we shall discuss (and beyond)
is in (\ref{W}). In particular, all the symmetries and quantum numbers of the
various parameters are already embodied in $W_{N_f,N_A}$.
The non-perturbative superpotential is derived in refs.~\cite{efgr,efgr2} by an
``integrating in'' procedure, following refs.~\cite{ILS,I}. The details can be
found in ref.~\cite{efgr2} and will not be presented here~\footnote{These were
discussed in a talk presented by E. Rabinovici in this meeting.}. Instead, in
the next sections, we list the main results concerning each of the theories,
$N_f,N_A,N_{3/2}$, case by case. Moreover, a few generalizations to other
gauge groups will be discussed.

\section{$b_1=6$: $N_f=N_A=N_{3/2}=0$}
\setcounter{equation}{0}
This is a pure $N=1$ supersymmetric $SU(2)$ gauge theory. The non-perturbative
effective superpotential is~\footnote{This can be read from eq. (\ref{W}) by
setting $\tr_{2N_f}(\cdot)=0$, $\det_{2N_f}(\cdot)=1$
(for example, $\pf X=1$, $\G=M$) when $N_f=0$, and $\tr_{N_A}(\cdot)=0$,
$\det_{N_A}(\cdot)=1$ (for example, $\det\Gamma =1$) when $N_A=0$; this will
also be used later.}
\beq
W_{0,0}=\pm 2 \L^3 .
\eeq{61}
The superpotential in eq. (\ref{61}) is non-zero due to gaugino (gluino)
condensation~\footnote{Recall that we regard $\L$ as a background chiral
superfield source.}. Let us consider gaugino condensation for general simple
groups~\cite{lec}. \\

\begin{center}
\bf{Pure $N=1$ Supersymmetric Yang-Mills Theories}
\end{center}
\vskip .1in

\noindent
Pure $N=1$ supersymmetric gauge theories are theories with pure superglue with
no matter. We consider a theory based on a simple group $G$. The theory
contains vector bosons $A_{\mu}$ and gauginos $\l_{\a}$ in the adjoint
representation of $G$. There is a classical $U(1)_R$ symmetry, gaugino number,
which is broken to a discrete $\Z_{2C_2}$ subgroup by instantons,
\beq
\langle (\l\l)^{C_2}\rangle = const. \L^{3C_2} ,
\eeq{llL3C}
where $C_2$ = the Casimir in the adjoint representation normalized such that,
for example, $C_2=N_c$ for $G=SU(N_c)$.

This theory confines, gets a mass gap, and there are $C_2$ vacua associated
with the spontaneous breaking of the $\Z_{2C_2}$ symmetry to $\Z_2$ by
gaugino condensation:
\beq
\langle \l\l \rangle = const. e^{2\pi i n/C_2} \L^3 , \quad n=1,...,C_2 .
\eeq{llL3}
Each of these $C_2$ vacua contributes $(-)^F=1$ and thus the Witten index is
${\rm Tr}(-)^F=C_2$. This physics is encoded in the generalization of eq.
(\ref{61}) to any $G$, giving
\beq
W_{eff}=e^{2\pi i n/C_2} C_2  \L^3 , \quad n=1,...,C_2 .
\eeq{WCL3}
For $G=SU(2)$ we have $C_2=2$. Indeed, the ``$\pm$'' in (\ref{61}), which
comes from the square-root appearing on the braces in (\ref{W}) when $b_1=6$,
corresponds, physically, to the two quantum vacua of a pure $N=1$
supersymmetric $SU(2)$ gauge theory.

The superpotentials (\ref{61}), (\ref{llL3}) can be derived by first adding
fundamental matter to pure $N=1$ supersymmetric YM theory (as we will do in the
next section), and then integrating it out.

\section{$b_1=5$: $N_f=1$, $N_A=N_{3/2}=0$}
\setcounter{equation}{0}
There is one case with $b_1=5$, namely, $SU(2)$ with one flavor. The
superpotential is
\beq
W_{1,0}={\L^5\over X}+mX ,
\eeq{51}
where $X$ and $m$ are defined by: $X_{ij}\equiv X\e_{ij}$,
$m_{ij}\equiv -m\e_{ij}$. The non-perturbative part of $W_{1,0}$ is proportional
to the one instanton action. The vacuum degeneracy of the classical low-energy
effective theory is lifted quantum mechanically; from eq. (\ref{51})
we see that, in the massless case, there is no vacuum at all.\\

\begin{center}
\bf{$SU(N_c)$ with $N_f<N_c$}
\end{center}
\vskip .1in

\noindent
Equation (\ref{51}) is a particular case of  $SU(N_c)$ with $N_f<N_c$
($N_f$ quarks $Q^i$ and $N_f$ anti-quarks $\Qb_{\ib}$,
$i,\ib=1,...,N_f$)~\cite{lec}. In these theories, by using holomorphy and
global symmetries,
\beq
\begin{array}{cccccc}
 &U(1)_Q &\times& U(1)_{\Qb} &\times& U(1)_R \\
Q :& 1 & & 0 & & 0 \\
\Qb :& 0 & & 1 & & 0 \\
\L^{3N_c-N_f} :& N_f & & N_f & & 2N_c-2N_f \\
W :& 0 & & 0 & & 2
\end{array}
\eeq{QQLW}
one finds that
\beq
W_{eff}=(N_c-N_f)\left({\L^{3N_c-N_f}\over \det X}\right)^{1\over N_c-N_f} ,
\eeq{WNflessNc}
where
\beq
X^i_{\, \ib} \equiv Q^i\Qb_{\ib}, \qquad i,\ib=1,...,N_f.
\eeq{Xiib}
Classically, $SU(N_c)$ with $N_f<N_c$ is broken down to $SU(N_c-N_f)$. The
effective superpotential in (\ref{WNflessNc}) is dynamically generated by
gaugino condensation in $SU(N_c-N_f)$ (for $N_f\leq N_c-2$)~\footnote{This is
the reason to the fractional $1/(N_c-N_f)$ power in (\ref{WNflessNc}) leading
to $C_2(SU(N_c-N_f))=N_c-N_f$ different phases in $W_{eff}$.}, and by
instantons (for $N_f=N_c-1$). \\

\begin{center}
\bf{The $SU(2)$ with $N_f=1$ Example}
\end{center}
\vskip .1in

\noindent
For example, let us elaborate on the derivation and physics of eq.
(\ref{51}). An $SU(2)$ effective theory with two doublets $Q_i^a$ has one
light degree of freedom: four $Q_i^a$ ($i=1,2$ is a flavor index, $a=1,2$ is a
color index; $2\times 2=4$) three out of which are eaten by $SU(2)$, leaving
$4-3=1$. This single light degree of freedom can be described by the gauge
singlet
\beq
X=Q_1Q_2 .
\eeq{XQ1Q2}
When $\langle X\rangle \neq 0$, $SU(2)$ is completely broken and, classically,
$W_{eff,class}=0$ (when $\langle X\rangle =0$ there are extra massless fields
due to an unbroken $SU(2)$). Therefore, the classical scalar
potential is identically zero. However, the one-instanton action is expected to
generate a non-perturbative superpotential.

The symmetries of the theory (at the classical level and with their
corresponding charges) are: $U(1)_{Q_1}$ = number of $Q_1$ fields (quarks or
squarks), $U(1)_{Q_2}$ = number of $Q_2$ fields (quarks or squarks), $U(1)_R$ =
$\{$number of gluinos$\}-\{$number of squarks$\}$. At the quantum level these
symmetries are anomalous -- $\partial_{\m} j^{\m} \sim F\tilde{F}$ -- and by
integrating both sides of this equation one gets a charge violation when there
is an instanton background $I$. The instanton background behaves like
\beq
I\sim e^{-8\pi^2/g^2(\m)}=\left({\L\over \m}\right)^{b_1} , \qquad b_1=5,
\eeq{Inst}
and it has four gluino zero-modes $\l$ and two squark zero-modes $q$.
Therefore, $\L^5$ has an $R$-charge = $4(\l)-2(q) = 2$,~\footnote{For $SU(N_c)$
with $N_f$ flavors, the instanton background $I$ has $2C_2=2N_c$ gluino
zero-modes $\l$ and $2N_f$ squark zero-modes $q$ and,
therefore, its $R$-charge is $R(I)$ =
number($\l$) $-$ number($q$) = $2N_c-2N_f$. Since $I\sim \L^{b_1}$ and
$b_1=3N_c-N_f$, we learn that $\L^{3N_c-N_f}$ has an $R$-charge = $2N_c-2N_f$,
as it appears in eq. (\ref{QQLW}).}
and we can summarize the quantum numbers of $X$ and $\L$:
\beq
\begin{array}{cccccc}
 &U(1)_{Q_1} &\times& U(1)_{Q_2} &\times& U(1)_R \\
X :& 1 & & 1 & & 0 \\
\L^5 :& 1 & & 1 & & 2
\end{array}
\eeq{XL5}
Moreover, ${\cal L}_{eff}=\int d^2\theta W_{eff}(X,\L)$ has charges = 0, while
$d^2\theta$ has charges:
\beq
\begin{array}{cccccc}
 &U(1)_{Q_1} &\times& U(1)_{Q_2} &\times& U(1)_R \\
d^2\theta :& 0 & & 0 & & -2
\end{array}
\eeq{d2tchar}
and, therefore, $W_{eff}$ has charges:
\beq
\begin{array}{cccccc}
 &U(1)_{Q_1} &\times& U(1)_{Q_2} &\times& U(1)_R \\
W_{eff} :& 0 & & 0 & & 2
\end{array}
\eeq{Wchar}
Finally, because $W_{eff}$ is holomorphic in $X,\L$, and is invariant under
symmetries, we must have
\beq
W_{eff}(X,\L)=c{\L^5\over X} .
\eeq{WXL}
This is an exact result. To determine the constant $c$, one needs to do the
one-instanton calculation. The calculation is well defined, and done by
Affleck, Dine and Seiberg~\cite{ads} with the result: $c\neq 0$. We can choose
units for $\L$ (or a renormalization scheme) to scale $c$ to 1. Therefore, as
was advocated, the vacuum degeneracy of the classical low-energy
effective theory is lifted quantum mechanically: the scalar potential is large
at strong coupling (small $|X|$) and goes asymptotically to 0 at large $|X|$
(as it should at weak coupling).

\section{$b_1=4$}
\setcounter{equation}{0}
There are two cases with $b_1=4$: either $N_f=2$, or $N_A=1$.
In both cases, the non-perturbative superpotential vanishes
and, in addition, there is a
constraint~\footnote{This is reflected in eq. (\ref{W}) by the vanishing of the
coefficient $(4-b_1)$ in front of the braces, leading to $W=0$, and the
singular power $1/(4-b_1)$ on the braces,  when $b_1=4$, which signals the
existence of a constraint.}.

\subsection{$N_f=2$, $N_A=N_{3/2}=0$}
The non-perturbative superpotential vanishes
\beq
W_{2,0}^{non-per.}=0 ,
\eeq{41}
and by the integrating in procedure we also get the quantum constraint:
\beq
\pf X=\L^4 .
\eeq{42}
At the classical limit, $\L\to 0$, the quantum constraint collapses into the
classical constraint, $\pf X=0$.\\

\begin{center}
\bf{$SU(N_c)$ with $N_f=N_c$}
\end{center}
\vskip .1in

\noindent
Equations (\ref{41}), (\ref{42}) are a particular case of  $SU(N_c)$ with
$N_f=N_c$.~\cite{lec} In these theories one obtains $W_{eff}=0$, and the
classical constraint $\det X - B\Bb=0$ is modified quantum mechanically to
\beq
\det X - B\Bb= \L^{2N_c} ,
\eeq{XBBL}
where
\ber
X^{i}_{\, \ib}&=&Q^i\Qb_{\ib} \quad ({\rm mesons}), \nonumber \\
B &=&\e_{i_1\dots i_{N_c}}Q^{i_1}\cdots Q^{i_{N_c}} \quad ({\rm baryon}),
\nonumber \\
\Bb &=&\e^{\ib_1\dots \ib_{N_c}}\Qb_{\ib_1}\cdots \Qb_{\ib_{N_c}} \quad
({\rm anti-baryon}).
\eer{XBBb}

\subsection{$N_f=0$, $N_A=1$, $N_{3/2}=0$}
The massless $N_A=1$ case is a pure $SU(2)$, $N=2$ supersymmetric Yang-Mills
theory. This model was considered in detail in ref.~\cite{SW1}.
The non-perturbative superpotential vanishes
\beq
W_{0,1}^{non-per.}=0 ,
\eeq{43}
and by the integrating in procedure we also get the quantum constraint:
\beq
M=\pm \L^2 .
\eeq{44}
This result can be understood because the starting point of the integrating
in procedure is a pure $N=1$ supersymmetric Yang-Mills theory. Therefore,
it leads us to the points at the verge of confinement in the moduli space.
These are the two singular points in the $M$ moduli space of the theory;
they are due to massless monopoles or dyons. Such excitations are not
constructed out of the elementary degrees of freedom and, therefore, there
is no trace for them in $W$. (This situation is different if $N_f\neq 0$,
$N_A=1$; in this case, monopoles are different manifestations of the
elementary degrees of freedom.)

\section{$b_1=3$}
\setcounter{equation}{0}
There are two cases with $b_1=3$: either $N_f=3$, or $N_A=N_f=1$. In both
cases, for vanishing bare parameters in (\ref{W}), the semi-classical limit,
$\L\to 0$, imposes the classical constraints, given by the equations of
motion: $\partial W=0$; however, quantum corrections remove the constraints.

\subsection{$N_f=3$, $N_A=N_{3/2}=0$}
The superpotential is
\beq
W_{3,0}=-{\pf X\over \L^3}+{1\over 2}\tr mX .
\eeq{31}
In the massless case, the equations $\partial_X W=0$ give the classical
constraints; in particular, the superpotential is proportional to a classical
constraint: $\pf X=0$. The negative power of $\L$, in eq. (\ref{31}) with
$m=0$, indicates that small values of $\L$ imply a semi-classical limit for
which the classical constraints are imposed. \\

\begin{center}
\bf{$SU(N_c)$ with $N_f=N_c+1$}
\end{center}
\vskip .1in

\noindent
Equation (\ref{31}) is a particular case of  $SU(N_c)$ with
$N_f=N_c+1$~\cite{lec}. In these theories one obtains
\beq
W_{eff}=-{\det X - X^i_{\, \ib} B_i \Bb^{\ib}\over \L^{2N_c-1}} ,
\eeq{WXBB}
where
\ber
X^{i}_{\, \ib}&=&Q^i\Qb_{\ib} \quad ({\rm mesons}), \nonumber \\
B_i &=& \e_{i j_1\dots j_{N_c}}Q^{j_1}\cdots Q^{j_{N_c}} \quad
({\rm baryons}), \nonumber \\
\Bb^{\ib}&=&\e^{\ib\jb_1\dots \jb_{N_c}}\Qb_{\jb_1}\cdots \Qb_{\jb_{N_c}}
\quad ({\rm anti-baryons}).
\eer{XBiBbi}
The equations $\partial_X W_{eff}=\partial_B W_{eff}=\partial_{\Bb} W_{eff}=0$
give the classical constraints:
\beq
\det X (X^{-1})_i^{\, \ib}-B_i\Bb^{\ib}=X_{\ib}^{\, i}B_i=X^i_{\,
\ib}\Bb^{\ib}=0 .
\eeq{XBXB0}
This is consistent with the negative power of $\L$ in $W_{eff}$ which implies
that in the semi-classical limit, $\L\to 0$, the classical constraints are
imposed.

\subsection{$N_f=1$, $N_A=1$, $N_{3/2}=0$}
In this case, the superpotential in (\ref{W}) reads
\beq
W_{1,1}=-{\pf X\over \L^3}\G^2+\tm M +{1\over 2}\tr mX + {1\over \sqrt{2}}\tr
\l Z .
\eeq{32}
Here $m$, $X$ are antisymmetric $2\times 2$ matrices, $\l$, $Z$ are
symmetric $2\times 2$ matrices and
\beq
\G=M+\tr (ZX^{-1})^2 .
\eeq{33}
This superpotential was found first in ref.~\cite{IS1}. To find the
quantum vacua, we solve the equations: $\partial_M W =\partial_X W
=\partial_Z W =0$. Let us discuss some properties of this theory:

\begin{itemize}

\item
The equations $\partial W=0$ can be re-organized into the singularity
conditions of an elliptic curve:
\beq
y^2=x^3+ax^2+bx+c
\eeq{34}
(and some other equations), where the coefficients $a,b,c$ are functions of
only the field $M$, the scale $\L$, the bare quark masses, $m$, and Yukawa
couplings, $\l$. Explicitly,
\beq
a=-M, \qquad b={\L^3\over 4}\pf m, \qquad c=-{\a\over 16} ,
\eeq{35}
where
\beq
\a={\L^6\over 4}\det\l .
\eeq{36}

\item
The parameter $x$, in the elliptic curve (\ref{34}), is given in terms of the
composite field:
\beq
x\equiv {1\over 2}\G .
\eeq{37}

\item
$W_{1,1}$ has $2+N_f=3$ vacua, namely, the  three singularities of the
elliptic curve in (\ref{34}), (\ref{35}). These are the three solutions,
$M(x)$, of the equations: $y^2=\partial y^2/\partial x=0$; the solutions
for $X$, $Z$ are given by the other equations of motion.

\item
The {\em 3 quantum vacua} are the vacua of the theory in the
{\em Higgs-confinement phase}.

\item
Phase transition points to the {\em Coulomb branch} are at
$X=0 \Leftrightarrow \tm=0$. Two of these singularities correspond to a
massless monopole or dyon, and are the quantum splitting of the classically
enhanced $SU(2)$ point. A third singularity is due to a massless quark; it is a
classical singularity: $M\sim m^2/\l^2$ for large $m$, and thus $M\to\infty$
when $m\to\infty$, leaving the two quantum singularities of the
$N_A=1$, $N_f=0$ theory.

\item
The elliptic curve defines the {\em effective Abelian coupling},
$\tau(M,\L,m,\l)$, in the Coulomb branch:\\

\begin{center}
\bf{Elliptic Curves and Effective Abelian Couplings}
\end{center}
\vskip .1in

\noindent
A torus can be described by the one complex dimensional curve in ${\bf C}^2$
$y^2=x^3+ax^2+bx+c$, where $(x,y)\in {\bf C}^2$ and $a,b,c$ are complex
parameters. The modular parameter of the torus is
\beq
\t (a,b,c)={\int_{\b} {dx \over y} \over \int_{\a} {dx \over y}} ,
\eeq{tab}
where $\a$ and $\b$ refer to a basis of cycles around the branch cuts of the
curve in the $x$ plan.

Alternatively, by redefining $x\to x-a/3$, we can take the curve into the
normal Weierstrass form
\beq
y^2=x^3+fx+g .
\eeq{yxfg}
In this form, the modular parameter $\t$ is determined (modulo $SL(2,\Z)$)
by the ratio $f^3/g^2$ through the relation
\beq
j(\t )={4(24f)^3\over 4f^3+27g^2} ,
\eeq{jtfg}
where $j$ is the modular-invariant holomorphic function~\cite{jfun}.
The singularities of the curve are located at the zeros of the
discriminant
\beq
\Delta=4f^3+27g^2.
\eeq{Dfg}

Therefore, eq. (\ref{34}) defines an $SL(2,\Z)$ section $\t$ in terms of $M$,
various coupling constants and a scale, which is singular at the zeroes of
$\Delta$.

\item
On the subspace of bare parameters, where the theory has an enhanced $N=2$
supersymmetry, the result in eq. (\ref{35}) coincides with the result
in~\cite{SW2} for $N_f=1$.

\item
In the massless case, there is a $Z_{4-N_f}=Z_3$ global symmetry acting on the
moduli space.

\item
When the masses and Yukawa couplings approach zero, all the 3 singularities
collapse to the origin. Such a point might be interpreted as a
new scale-invariant theory~\cite{lec}. As before, the
negative power of $\L$, in eq. (\ref{32}) with $\tm=m=\l=0$,
indicates that small values of
$\L$ imply a semi-classical limit for which the classical constraint, $\G=0$,
is imposed. Indeed, for vanishing bare parameters, the equations of motion
are solved by any $M,X,Z$ obeying $\G=0$.

\item
By tuning the quark mass $m$, one can find a situation where two out of the
three singularities degenerate~\cite{apsw}. At the phase transition to the
Coulomb branch, this theory has massless, mutually non-local charged
degrees of freedom; it is a new non-trivial, interacting $N=1$ SCFT.

\end{itemize}

\section{$b_1=2$}
\setcounter{equation}{0}
There are three cases with $b_1=2$: $N_f=4$, or $N_A=1$, $N_f=2$, or $N_A=2$.
In all three cases, for vanishing bare parameters in (\ref{W}), there
are extra massless degrees of freedom not included in the procedure; those are
expected due to a non-Abelian conformal theory.

\subsection{$N_f=4$, $N_A=N_{3/2}=0$}
The superpotential is
\beq
W_{4,0}=-2{(\pf X)^{{1\over 2}}\over \L}+{1\over 2}\tr mX .
\eeq{21}
This theory is a particular case of $SU(N_c)$ with $N_f>N_c+1$, considered
in ref.~\cite{lec}.
In the massless case, the superpotential is proportional to the square-root of
a classical constraint: $\pf X=0$.  The branch cut at $\pf X=0$ signals the
appearance of extra massless degrees of freedom at these points.
Therefore, we make use of the superpotential only in the presence of masses,
$m$, which fix the vacua away from such points. \\

\begin{center}
\bf{Non-Abelian Superconformal Field Theories and Duality}
\end{center}
\vskip .1in

\noindent
To describe this phenomenon of appearance of extra interacting, massless degrees
of freedom at $\langle X\rangle =0$ in eq. (\ref{21}), we shall discuss briefly
$SU(N_c)$ with $N_f$ flavors IR theories, which are non-trivial $4d$
superconformal field theories (SCFTs), and the Seiberg duality~\cite{S2}
which ``predicts,'' in particular, the appearance of interacting, massless
degrees of freedoms at the points where the superpotential develops a branch
cut.

\begin{itemize}
\item
{\em Non-Abelian Coulomb Phase}: there is a strong evidence that for $SU(N_c)$
with ${3\over 2}N_c<N_f<3N_c$ the theory is in an interacting, non-Abelian
Coulomb phase (in the IR and for $m=0$). In this range of $N_f$ the theory is
asymptotically free. Namely, at short distance the coupling constant $g$ is
small, and it becomes larger at larger distance. However, it is argued that for
${3\over 2}N_c<N_f<3N_c$, $g$ does not grow to infinity, but it reaches a
finite value -- a fixed point of the renormalization group.
Therefore, for ${3\over 2}N_c<N_f<3N_c$, the IR theory is a non-trivial $4d$
SCFT. The elementary quarks and gluons are not confined but appear as
interacting massless particles. The potential between external massless
electric sources behaves as $V\sim 1/R$, and thus one refers to this phase of
the theory as the non-Abelian Coulomb phase.
\item
{\em The Seiberg Duality}: it is claimed~\cite{S2} that in the IR an $SU(N_c)$
theory with $N_f$ flavors is dual to $SU(N_f-N_c)$ with $N_f$ flavors but, in
addition to dual quarks, one should also include interacting, massless scalars.
This is the origin to the branch cut in $W_{eff}$ at $\langle X\rangle =0$,
because $W_{eff}$ does not include these light modes which must appear at
$\langle X\rangle =0$.

The quantum numbers of the quarks and anti-quarks of the $SU(N_c)$ theory
with $N_f$ flavors (= theory A) are~\footnote{Here we use the convention where
the axial $U(1)_A$ symmetry is anomalous while the $R$-charge is conserved;
previously, we used the opposite convention.} \\

\begin{center}
{\bf A. $SU(N_c)$, $N_f$: The Electric Theory}
\end{center}

\beq
\begin{array}{cccccccc}
 &SU(N_f)_L &\times& SU(N_f)_R &\times& U(1)_B &\times& U(1)_R \\
Q :& N_f& &1 & & 1 & & 1-{N_c\over N_f} \\
\Qb : & 1& & {\bar N_f} & & -1 & & 1-{N_c\over N_f}
\end{array}
\eeq{ABR}
The quantum numbers of the dual quarks $q_i$ and anti-quarks $\bar{q}^{\ib}$
of the $SU(N_f-N_c)$ theory with  $N_f$ flavors theory (= theory B) and its
massless scalars $X^i_{\, \ib}$ are \\

\begin{center}
{\bf B. $SU(N_f-N_c)$, $N_f$: The Magnetic Theory}
\end{center}

\beq
\begin{array}{cccccccc}
 &SU(N_f)_L &\times& SU(N_f)_R &\times& U(1)_B &\times& U(1)_R \\
q :& {\bar N_f}& &1 & & {N_c\over N_f-N_c} & & {N_c\over N_f} \\
\bar{q} : & 1& & N_f & & -{N_c\over N_f-N_c} & & {N_c\over N_f} \\
X: &N_f & & {\bar N_f} & & 0 & & 2\left(1-{N_c\over N_f}\right)
\end{array}
\eeq{BBR}

The $U(1)_R$ numbers in both theory A and theory B are the non-anomalous
$R$-charges. The baryon numbers (under $U(1)_B$) of the dual quarks and
anti-quarks are designed to construct dual baryons and and anti-baryons with
$B$-charge equal to the $B$-charge of the baryons and anti-baryons in theory A.
We notice that these baryon numbers are fractional and, therefore,
the dual quarks cannot be represented by a polynomial in the original quarks
$Q$ and antiquarks $\Qb$; in terms of theory A, they are some (non-local,
composite) magnetic degrees of freedom. Thus, while we refer to theory A as
the ``electric theory,'' we call its dual the ``magnetic theory,'' and the
duality relating theory A to theory B is an electric-magnetic duality.

The massless scalars in theory B are independent degrees of freedom
($X\neq q\bar{q}$ except for the self-dual case: $N_f=2N_c$). Moreover, one
should add the tree-level superpotential
\beq
W_{tree}=X^i_{\, \ib} q_i \bar{q}^{\ib} ,
\eeq{WXqqb}
and, therefore, these massless scalars are interacting, as advocated.
\item
{\em Non-Trivial Checks of Duality}:
\begin{enumerate}
\item
{\em `t Hooft Anomaly Matching Condition}: it says that if two theories are
equivalent they must have the same {\em global} symmetries and, therefore, if
we make this global symmetry local in both theories we must obtain the same
anomaly numbers with respect to such a gauge symmetry. It was checked~\cite{S2}
that, miraculously, both theory A and theory B have the same anomalies:
\beq
\begin{array}{ll}
U(1)_B^3 : & 0 \\
U(1)_B U(1)_R^2 : & 0 \\
U(1)_B^2 U(1)_R : & -2N_c^2 \\
SU(N_f)^3 : & N_c d^{(3)}(N_f) \\
SU(N_f)^2 U(1)_R : & -{N_c^2\over N_f} d^{(2)}(N_f) \\
SU(N_f)^2 U(1)_B : & N_c d^{(2)}(N_f) \\
U(1)_R : & -N_c^2-1 \\
U(1)_R^3 : & N_c^2-1-2{N_c^4\over N_f^2}
\end{array}
\eeq{UUU}
Here $d^{(3)}(N_f)={\rm Tr} T_f^3$ of the global $SU(N_f)$ symmetries, where
$T_f$ are generators in the fundamental representation, and
$d^{(2)}(N_f)={\rm Tr} T_f^2$
\item
{\em Deformations}: theory A and theory B have the same quantum moduli space
                    of deformations.
\end{enumerate}
\end{itemize}

\begin{center}
\bf
Remarks
\end{center}

\begin{itemize}
\item
Electric-magnetic duality exchanges strong coupling with weak coupling (this
can be read off from the beta-functions), and it interchanges a theory in the
Higgs phase with a theory in the confining phase.
\item
Strong-weak coupling duality also relates an $SU(N_c)$ theory with $N_f\geq
3N_c$ to an $SU(N_f-N_c)$ theory. $SU(N_c)$ with $N_f\geq 3N_c$ is in a
non-Abelian {\em free electric phase}: in this range the theory is not
asymptotically free. Namely, because of screening, the coupling constant
becomes smaller at large distance. Therefore, the spectrum of the theory at
large distance can be read off from the Lagrangian -- it consists of the
elementary quarks and gluons. The long distance behavior of the potential
between external electric test charges is
\beq
V(R) \sim {1 \over R\log (R\L)} \sim {e^2(R)\over R}, \qquad
e(R\to\infty)\to 0.
\eeq{eRlogR}
For $N_f\geq 3N_c$, the theory is thus in a non-Abelian free electric phase;
the massless electrically charged fields renormalize the charge to zero at long
distance as $e^{-2}(R)\sim \log (R\L)$.
The potential of magnetic test charges behave at large distance $R$ as
\beq
V(R) \sim {\log (R\L) \over R} \sim {m^2(R)\over R}, \quad \implies \quad
e(R)m(R)\sim 1 .
\eeq{mRlogR}

$SU(N_c)$ with $N_f\geq 3N_c$ is dual to $SU(\tilde{N}_c)$ with
$\tilde{N}_c +2 \leq N_f \leq {3\over 2}\tilde{N}_c$, where
$\tilde{N}_c =N_f-N_c$. This dual theory is in a non-Abelian {\em free magnetic
phase}: there are massless magnetically charged fields $X,q,\bar{q}$ and a
potential between external electric test charges with a conjectured behavior at
large distance as
\beq
V(R) \sim {\log (R\L) \over R} \sim {e^2(R)\over R}, \qquad
e(R\to\infty)\to \infty .
\eeq{eRlogRd}
For $\tilde{N}_c +2 \leq N_f \leq {3\over 2}\tilde{N}_c$, the massless magnetic
monopoles renormalize the electric coupling constant to infinity at large
distance, with a conjectured behavior $e^2(R)\sim \log (R\L)$.
The potential of magnetic test charges behaves at large distance $R$ as
\beq
V(R) \sim {1 \over R\log (R\L)}\sim {m^2(R) \over R} \quad \implies \quad
e(R)m(R)\sim 1 .
\eeq{mRlogRd}
\item
The Seiberg duality can be generalized in many other cases, including a
variety of matter supermultiplets (like superfields in the adjoint
representation~\cite{K}) and other gauge groups~\cite{moreduality}.
\end{itemize}

\subsection{$N_f=2$, $N_A=1$, $N_{3/2}=0$}
In this case, the superpotential in (\ref{W}) reads
\beq
W_{2,1}=-2{(\pf X)^{{1\over 2}}\over \L}\G
+\tm M +{1\over 2}\tr mX + {1\over \sqrt{2}}\tr \l Z .
\eeq{22}
Here $m$, $X$ are antisymmetric $4\times 4$ matrices, $\l$, $Z$ are
symmetric $4\times 4$ matrices and $\G$ is given in eq. (\ref{33}).
As in section 7.2, to find the quantum vacua, we solve the equations:
$\partial W =0$. Let us discuss some properties of this theory:

\begin{itemize}

\item
The equations $\partial W=0$ can be re-organized into the singularity
conditions of an elliptic curve (\ref{34}) (and some other equations), where
the coefficients $a,b,c$ are functions of
only the field $M$, the scale $\L$, the bare quark masses, $m$, and Yukawa
couplings, $\l$. Explicitly~\cite{efgr,efgr2},
\beq
a= -M, \qquad b= -{\a\over 4}+{\L^2\over 4}\pf m , \qquad
c= {\a\over 8}\Big(2M+\tr(\m^2)\Big),
\eeq{23}
where
\beq
\a={\L^4\over 16}\det\l , \qquad \m=\l^{-1}m .
\eeq{24}

\item
As in section 7.2, the parameter $x$, in the elliptic curve (\ref{34}),
is given in terms of the composite field:
\beq
x\equiv {1\over 2}\G .
\eeq{25}
Therefore, we have identified a physical meaning of the parameter $x$.

\item
$W_{2,1}$ has $2+N_f=4$ vacua, namely, the four singularities of the
elliptic curve in (\ref{34}), (\ref{23}). These are the four solutions,
$M(x)$, of the equations: $y^2=\partial y^2/\partial x=0$; the solutions
for $X$, $Z$ are given by the other equations of motion.

\item
The {\em 4 quantum vacua} are the vacua of the theory in the
{\em Higgs-confinement phase}.

\item
Phase transition points to the {\em Coulomb branch} are at
$X=0 \implies \tm=0$.
Therefore, we conclude that the elliptic curve defines the {\em effective
Abelian coupling}, $\tau(M,\L,m,\l)$, in the Coulomb branch.

\item
On the subspace of bare parameters, where the theory has an enhanced $N=2$
supersymmetry, the result in eq. (\ref{23}) coincides with the result
in~\cite{SW2} for $N_f=2$.

\item
In the massless case, there is a $Z_{4-N_f}=Z_2$ global symmetry acting on the
moduli space.

\item
As in section 7.2, the negative power of $\L$, in eq. (\ref{22}) with
$\tm=m=\l=0$, indicates that small values of $\L$ imply a semi-classical
limit for which the classical constraints are imposed. Indeed, for vanishing
bare parameters, the equations $\partial W=0$ are equivalent to the classical
constraints, and their solutions span the Higgs moduli space~\cite{efgr4}.

\item
For special values of the bare masses and Yukawa couplings, some of the 4 vacua
degenerate. In some cases, it may lead to points where mutually non-local
degrees of freedom are massless, similar to the situation in pure $N=2$
supersymmetric gauge theories, considered in~\cite{AD}. For example, when the
masses and Yukawa couplings approach zero, all the 4 singularities
collapse to the origin. Such points might be interpreted as in a
{\em non-Abelian Coulomb phase}~\cite{lec} or new non-trivial, interacting,
$N=1$ SCFTs.

\item
The singularity at $X=0$ (in $\G$) and the branch cut at $\pf X=0$
(due to the $1/2$ power in eq. (\ref{22}))
signal the appearance of extra massless degrees of freedom at these points;
those are expected similar to references~\cite{S2,K}.
Therefore, we make use of the superpotential only in the presence of bare
parameters, which fix the vacua away from such points.

\end{itemize}

\subsection{$N_f=0$, $N_A=2$, $N_{3/2}=0$}
In this case, the superpotential in eq. (\ref{W}) reads
\beq
W_{0,2}=\pm 2{\det M\over \L}+\tr \tm M .
\eeq{26}
Here $\tm$, $M$ are $2\times 2$ symmetric matrices, and the ``$\pm$'' comes
from the square-root, appearing on the braces in (\ref{W}), when $b_1=2$.
The superpotential in eq. (\ref{26}) is the one presented in~\cite{IS1,IS2}
on the confinement and the oblique confinement branches~\footnote{
The fractional power $1/(4-b_1)$ on the braces
in (\ref{W}), for any theory with $b_1\leq 2$, may indicate a similar
phenomenon, namely, the existence of confinement and oblique
confinement branches of the theory, corresponding to the $4-b_1$
phases due to the fractional power. It is plausible that, for $SU(2)$, such
branches are related by a discrete symmetry.}
(they are related by a discrete symmetry~\cite{lec}). This theory has
{\em two quantum vacua}; these become the phase transition points to the
Coulomb branch when $\det \tm =0$. The moduli space may also contain a
non-Abelian Coulomb phase when the two singularities degenerate at the
point $M=0$~\cite{IS1}; this happens when $\tm=0$.
At this point, the theory has
extra massless degrees of freedom and, therefore, $W_{0,2}$ fails to
describe the physics at $\tm=0$. Moreover, at $\tm=0$, the theory has other
descriptions via an electric-magnetic triality~\cite{lec}.

\section{$b_1=1$}
\setcounter{equation}{0}
There are four cases with $b_1=1$: $N_f=5$, or $N_A=1$, $N_f=3$, or $N_A=2$,
$N_f=1$, or $N_{3/2}=1$.

\subsection{$N_f=5$, $N_A=N_{3/2}=0$}
The superpotential is
\beq
W_{5,0}=-3{(\pf X)^{{1\over 3}}\over \L^{{1\over 3}}}+{1\over 2}\tr mX .
\eeq{11}
This theory is a particular
case of $SU(N_c)$ with $N_f>N_c+1$. The discussion in section 8.1 is
relevant in this case too.

\subsection{$N_f=3$, $N_A=1$, $N_{3/2}=0$}
In this case, the superpotential in (\ref{W}) reads
\beq
W_{3,1}=-3{(\pf X)^{{1\over 3}}\over \L^{{1\over 3}}}\G^{{2\over 3}}
+\tm M +{1\over 2}\tr mX + {1\over \sqrt{2}}\tr \l Z .
\eeq{12}
Here $m$, $X$ are antisymmetric $6\times 6$ matrices, $\l$, $Z$ are
symmetric $6\times 6$ matrices and $\G$ is given in eq. (\ref{33}).
Let us discuss some properties of this theory:

\begin{itemize}

\item
The equations $\partial W=0$ can be re-organized into the singularity
conditions of an elliptic curve (\ref{34}) (and some other equations), where
the coefficients $a,b,c$ are~\cite{efgr,efgr2}
\ber
a&=& -M-\a , \qquad b\,\, =\,\, 2\a  M +
{\a\over 2}\tr(\m^2) + {\L\over 4} \pf m , \nonumber\\
c&=&{\a\over 8}\Big( -8M^2-4M \tr(\m^2) - [\tr(\m^2)]^2 + 2\tr(\m^4) \Big) ,
\eer{13}
where
\beq
\a={\L^2\over 64}\det\l , \qquad \m=\l^{-1}m .
\eeq{14}
In eq. (\ref{13}) we have shifted the quantum field $M$ to
\beq
M\to M-\a/4 .
\eeq{15}

\item
The parameter $x$, in the elliptic curve (\ref{34}), is given in terms of the
composite field:
\beq
x\equiv {1\over 2}\G + {\a\over 2}.
\eeq{16}
Therefore, as before, we have identified a physical meaning of the parameter
$x$.

\item
$W_{3,1}$ has $2+N_f=5$ {\em quantum vacua}, corresponding to the 5
singularities of the elliptic curve (\ref{34}), (\ref{13});
these are the vacua of the theory in the {\em Higgs-confinement phase}.

\item
{}From the phase transition points to the {\em Coulomb branch}, we conclude
that
the elliptic curve defines the {\em effective Abelian coupling},
$\tau(M,\L,m,\l)$, for arbitrary bare masses and Yukawa couplings.
As before, on the subspace of bare parameters, where the theory has $N=2$
supersymmetry, the result in eq. (\ref{13}) coincides with the result
in~\cite{SW2} for $N_f=3$.

\item
For special values of the bare masses and Yukawa couplings, some of the 5 vacua
degenerate. In some cases, it may lead to points where mutually non-local
degrees of freedom are massless, and might be interpreted as in a
{\em non-Abelian Coulomb phase} or another
new {\em superconformal theory} in four
dimensions (see the discussion in sections 7.2 and 8.2).

\item
The singularity  and branch cuts in $W_{3,1}$
signal the appearance of extra massless degrees of freedom at these points.

\item
The discussion in the end of sections 7.2 and 8.2 is relevant here too.

\end{itemize}

\subsection{$N_f=1$, $N_A=2$, $N_{3/2}=0$}
In this case, the superpotential in (\ref{W}) reads~\cite{efgr2}
\beq
W_{1,2}=-3{(\pf X)^{1/3}\over \L^{1/3}}(\det\G)^{2/3}+\tr \tm M
+{1\over 2}\tr mX + {1\over \sqrt{2}} \tr \l^{\a} Z_{\a}.
\eeq{17}
Here $m$ and $X$ are  antisymmetric $2\times 2$ matrices,
$\l^{\a}$ and $Z_{\a}$ are symmetric $2\times 2$ matrices, $\a=1,2$,
$\tm$, $M$ are $2\times 2$ symmetric matrices and $\G_{\a\b}$ is given in eq.
(\ref{G}).
This theory has {\em 3 quantum vacua} in the Higgs-confinement branch. At the
phase transition points to the Coulomb branch, namely, when $\det\tm =0
\Leftrightarrow \det M=0$, the equations of motion can be re-organized into the
singularity conditions of an elliptic curve (\ref{34}). Explicitly, when
$\tm_{22}=\tm_{12}=0$, the coefficients $a,b,c$ in (\ref{34}) are~\cite{efgr2}
\beq
a=-M_{22},\qquad b={\L\tm_{11}^2\over 16}\pf m,\qquad
c=-\Big({\L\tm_{11}^2\over 32}\Big)^2 \det\l_2 .
\eeq{18}
However, unlike the $N_A=1$ cases, the equations $\partial W=0$ {\em cannot} be
re-organized into the singularity condition of an elliptic curve, in general.
This result makes sense, physically, since an elliptic curve is expected to
``show up'' only at the phase transition points to the Coulomb branch.
For special values of the bare parameters, there are points in the moduli
space where (some of) the singularities degenerate; such points might be
interpreted as in a non-Abelian Coulomb phase, or new superconformal theories.
For more details, see ref.~\cite{efgr2}.

\subsection{$N_f=N_A=0$, $N_{3/2}=1$}
This chiral theory was shown to have
$W_{0,0}^{non-per.}(N_{3/2}=1)=0$;~\cite{ISS} perturbing it by a tree-level
superpotential, $W_{tree}=gU$, where $U$ is given in (\ref{XMZ}), may
lead to dynamical supersymmetry breaking~\cite{ISS}.

\section{$b_1=0$}
\setcounter{equation}{0}
There are five cases with $b_1=0$: $N_f=6$, or $N_A=1$, $N_f=4$, or
$N_A=N_f=2$, or $N_A=3$, or $N_{3/2}=N_f=1$. These theories have vanishing
one-loop beta-functions in either conformal or infra-red free beta-functions
and, therefore, will possess extra structure.

\subsection{$N_f=6$, $N_A=N_{3/2}=0$}
This theory is a particular case of $SU(N_c)$ with $N_f=3N_c$; the electric
theory is free in the infra-red~\cite{lec}.~\footnote{
A related fact is that
(unlike the $N_A=1$, $N_f=4$ case, considered next) in the (would be)
superpotential, $W_{6,0}=-4\L^{-b_1/4}(\pf X)^{1/4}+{1\over 2}\tr mX$,
it is impossible to construct the matching ``$\L^{b_1}$''$\equiv f(\t_0)$,
where $\t_0$ is the non-Abelian gauge coupling constant,
in a way that respects the global symmetries.}

\subsection{$N_f=4$, $N_A=1$, $N_{3/2}=0$}
In this case, the superpotential in (\ref{W}) reads
\beq
W_{4,1}=-4{(\pf X)^{{1\over 4}}\over \L^{{b_1\over 4}}}\G^{{1\over 2}}
+\tm M +{1\over 2}\tr mX + {1\over \sqrt{2}}\tr \l Z .
\eeq{01}
Here $m$, $X$ are antisymmetric $8\times 8$ matrices, $\l$, $Z$ are
symmetric $8\times 8$ matrices, $\G$ is given in eq. (\ref{33}) and
\beq
\L^{b_1}\equiv 16\a(\t_0)^{1/2}(\det \l)^{-1/2} ,
\eeq{02}
where $\a(\t_0)$ will be presented in eq. (\ref{04}).
Let us discuss some properties of this theory:

\begin{itemize}

\item
The equations $\partial W=0$ can be re-organized into the singularity
conditions of an elliptic curve (\ref{34}) (and some other equations), where
the coefficients $a,b,c$ are~\cite{efgr,efgr2}
\ber
a&=&{1\over \b^2}\Big\{
2{\a+1\over \a-1}M+{8\over \b^2}{\a\over (\a-1)^2}\tr(\m^2)\Big\}, \nonumber\\
b&=&{1\over \b^4}\Big\{
-16{\a\over (\a-1)^2} M^2 + {32\over \b^2}{\a(\a+1)\over (\a-1)^3}M\tr(\m^2)
\nonumber\\
&-&{8\over \b^4}{\a\over (\a-1)^2}\Big[(\tr (\m^2))^2-2\tr(\m^4)\Big]+
{4\over \b^4}{(\a+1)\L^{b_1}\over (\a-1)^2} \pf m\Big\} , \nonumber\\
c&=& {1\over \b^6}\Big\{
-32{\a(\a+1)\over (\a-1)^3}M^3+{32\over \b^2}{\a(\a+1)^2\over
(\a-1)^4}M^2\tr(\m^2)\nonumber\\
&+&M\Big[-{16\over \b^4}{\a(\a+1)\over (\a-1)^3}
\Big((\tr(\m^2))^2-2\tr(\m^4)\Big) + {32\over \b^4}{\a\L^{b_1}\over (\a-1)^3}
\pf m \Big]\nonumber\\
&-&{32\over \b^6}{\a\over (\a-1)^2}\Big[\tr(\m^2)\tr(\m^4)-{1\over
6}(\tr(\m^2))^3-{4\over 3}\tr(\m^6)\Big]\Big\}.
\eer{03}
Here $\m=\l^{-1}m$ and $\a$, $\b$ are functions of $\tau_0$, the non-Abelian
gauge coupling constant; comparison with ref.~\cite{SW2} gives
\beq
\a(\tau_0)\equiv {``\L^{2b_1}"\over 256} \det\l
=\left({\th_2^2-\th_3^2\over \th_2^2+\th_3^2}\right)^2, \qquad
\b(\tau_0)={\sqrt{2}\over \th_2\th_3},
\eeq{04}
where
\beq
\th_2(\tau_0)=\sum_{n\in Z}(-1)^n e^{\pi i \tau_0 n^2}, \qquad
\th_3(\tau_0)=\sum_{n\in Z}e^{\pi i \tau_0 n^2}, \qquad
\tau_0={\th_0\over \pi}+{8\pi i\over g_0^2}.
\eeq{05}
In eq. (\ref{03}) we have shifted the quantum field $M$ to
\beq
M\to \b^2 M-\a\tr\m^2/(\a-1) .
\eeq{06}

\item
The parameter $x$, in the elliptic curve (\ref{34}), is given in terms of
the composite field:
\beq
x\equiv {1\over \b^4}\Big[\G - {4\a\over (\a-1)^2}\tr\m^2\Big].
\eeq{07}

\item
$W_{4,1}$ has $2+N_f=6$ {\em quantum vacua}, corresponding to the 6
singularities of the elliptic curve (\ref{34}), (\ref{03}); these are the
vacua of the theory in the {\em Higgs-confinement phase}.

\item
As before, from the phase transition points to the {\em Coulomb branch}, we
conclude that the elliptic curve defines the {\em effective Abelian coupling},
$\tau(M,\L,m,\l)$, for arbitrary bare masses and Yukawa couplings.
On the subspace of bare parameters, where the theory has $N=2$
supersymmetry, the result in eq. (\ref{03}) coincides with the result
in~\cite{SW2} for $N_f=4$.

\item
The discussion in the end of sections 7.2, 8.2 and 9.2 is relevant
here too.

\item
We can get the results for $N_A=1$, $N_f<4$, by integrating out flavors.

\item
In all the $N_A=1$, $N_f\neq 0$ cases we {\em derived} the result that $\t$ is
a section of an $SL(2,\Z)$ bundle over the moduli space and over the parameters
space of bare masses and Yukawa couplings (since $\t$ is a modular parameter of
a torus).

\end{itemize}

\subsection{$N_f=2$, $N_A=2$, $N_{3/2}=0$}
It was argued that this theory is infra-red free~\cite{efgr2}.~\footnote{A
related fact is that (unlike the $N_A=1$, $N_f=4$ case) it is impossible to
construct a matching, ``$\L^{b_1}$''$\equiv\a(\t_0)f(\l^{\a})$, in a way
that respects the global symmetries.}

\subsection{$N_f=0$, $N_A=3$, $N_{3/2}=0$}
In this case, the superpotential in eq. (\ref{W}) reads
\beq
W_{0,3}=-4{(\det M)^{{1\over 2}}\over \L^{{b_1\over 4}}}+\tr \tm M .
\eeq{08}
Here $\tm$, $M$ are $3\times 3$ symmetric matrices. The superpotential
(\ref{08}) equals to the  tree-level superpotential, $W_{tree}=\l\det\P$,
where, schematically,  $\det\P\sim \e\P\P\P\sim (\det M)^{1/2}$ is the
(antisymmetric) coupling of the three triplets, $\P_{\a}$.
This result coincides
with the one derived in~\cite{IS2}. Therefore, we identify the matching
``$\L^{-b_1/4}$''$\equiv \l f(\t_0)$, which respects the global symmetries.
In the massless case, this theory flows to an $N=4$ supersymmetric Yang-Mills
fixed point.

\subsection{$N_f=1$, $N_A=0$, $N_{3/2}=1$}
It was argued that this theory is infra-red free~\cite{efgr}.

\section{More Results}
\setcounter{equation}{0}

We have summarized some old and new results in $N=1$ supersymmetric $SU(2)$
gauge theories, and generalizations to other groups. More new results,
which were derived in~\cite{efgr2,efgr3,efgr4} along the lines of section
3, were presented in this meeting by E. Rabinovici. In particular, in
ref.~\cite{efgr3}, we discussed how the structure of massless monopoles in
supersymmetric theories with a Coulomb phase can be obtained from effective
superpotentials for a phase with a confined photon. The technique was
illustrated for $SU(N_c)$ in~\cite{efgr3}, and results for other gauge groups
were reported this meeting and will appear in~\cite{efgr4}.

Finally, let us remark on the interplay between the properties of
supersymmetric gauge theories and the properties of
strings and extended objects. Much progress was made in understanding gauge
dynamics by applying some ``string theory intuition,'' by taking the
large Planck-mass limit, or by using probes in non-perturbative string
backgrounds. Conversely, much progress was made in understanding
non-perturbative string dynamics by applying some ``gauge theory
intuition.'' The field develops fast, and an updated list of references would
be enormous and will require a new update very soon.

\section*{Acknowledgments}
This work is supported in part by BSF -- American-Israel Bi-National
Science Foundation, and by the Israel Science Foundation founded by the
Israel Academy of Sciences and Humanities -- Centers of Excellence Program.


\begin{thebibliography}{6666}
\bibitem{moreduality} A partial list of refs. can be found in~\cite{lec},
               however, the field develops fast, and an updated list would
               be enormous and will require a new update very soon.
\end{thebibliography}
\end{document}